\newif\ifonecol
\onecoltrue

\ifonecol
    \newcommand{\CLASSINPUTbaselinestretch}{1.6}
    \documentclass[12pt,draftclsnofoot,onecolumn]{IEEEtran}
\else
    \documentclass[lettersize,journal]{IEEEtran}
\fi

\usepackage{amsmath,amsfonts, amssymb}
\usepackage{array}
\usepackage[caption=false,font=normalsize,labelfont=sf,textfont=sf]{subfig}
\usepackage{textcomp}
\usepackage{stfloats}
\usepackage{url}
\usepackage{verbatim}
\usepackage{graphicx}
\usepackage{xcolor}
\usepackage{balance}
\usepackage{algorithm}
\usepackage{algorithmicx}
\usepackage{algpseudocode}
\usepackage{booktabs}
\usepackage{adjustbox}
\usepackage{bm}

\DeclareMathOperator*{\argmax}{arg\,max}

\usepackage[framemethod=tikz]{mdframed}
\usepackage{soul}
\usepackage{color}

\soulregister\cite7
\soulregister\ref7
\soulregister\eqref7
\soulregister\pageref7

\ifonecol
\else
    \def\BibTeX{{\rm B\kern-.05em{\sc i\kern-.025em b}\kern-.08em
    T\kern-.1667em\lower.7ex\hbox{E}\kern-.125emX}}
\fi
\begin{document}
\title{Joint Spectrum Sensing and Resource Allocation for OFDMA-based Underwater Acoustic Communications}
\author{Minwoo Kim, Youngchol Choi, Yeongjun Kim, Eojin Seo, and Hyun Jong Yang,~\IEEEmembership{Member,~IEEE}

}



\maketitle

\begin{abstract}
Underwater acoustic (UWA) communications generally rely on cognitive radio (CR)-based ad-hoc networks due to challenges such as long propagation delay, limited channel resources, and high attenuation.
To address the constraints of limited frequency resources, UWA communications have recently incorporated orthogonal frequency division multiple access (OFDMA), significantly enhancing spectral efficiency (SE) through multiplexing gains.
Still, {the} low propagation speed of UWA signals, combined with {the} dynamic underwater environment, creates asynchrony in multiple access scenarios.
This causes inaccurate spectrum sensing as inter-carrier interference (ICI) increases, which leads to difficulties in resource allocation.
As efficient resource allocation is essential for achieving high-quality communication in OFDMA-based CR networks, these challenges degrade communication reliability in UWA systems.
To resolve the issue, we propose an end-to-end sensing and resource optimization method using deep reinforcement learning (DRL) in an OFDMA-based UWA-CR network.
Through extensive simulations, we confirm that the proposed method is superior to baseline schemes, outperforming other methods by 42.9 \% in SE and 4.4 \% in communication success rate.
\end{abstract}


\section{Introduction}  \label{sec:introduction}

\IEEEPARstart{U}{nlike}
terrestrial network communications that rely on electromagnetic waves, underwater communications utilize acoustic waves {due to the high attenuation of electromagnetic waves in water}.
However, underwater acoustic (UWA) communications suffer from slow propagation speed of acoustic waves, limited frequency resources, severe multipath spread, and rapid time variation, which pose significant obstacles to reliable communications \cite{DOMINGO2008163}.

To solve these issues, efforts such as the adoption of orthogonal frequency division multiplexing (OFDM) waveform, and the use of cognitive radio (CR) have been made \cite{10266670}.
Specifically, OFDM allows the receiver to deal with long multipath spread of UWA channels, and CR-based ad-hoc network allows for an effective use of the frequency resources.
Despite such efforts to achieve reliable UWA communications, there are still problems to overcome.

\subsection{Challenges of UWA Communications}  \label{subsec:challenges}

A significant challenge in UWA communications arises from the low propagation speed of acoustic waves.
{The} slow propagation {speed} makes synchronization between users difficult in dynamic underwater environment.
In UWA-CR networks, synchronization using the control channel is impractical due to the substantial delay {in} the control information round-trip time, which can {extend} to {several} tens of seconds.
Similarly, significant arrival time difference{s} of beacons across different nodes make beacon-based synchronization inefficient.
Therefore, {unlike in aerial communications \cite{9422153, 9293155}}, asynchrony is inevitable in multiple access UWA-CR networks, leading to high levels of inter-carrier interference (ICI) \cite{1200412}.
This can lead to inaccurate sensing, where unused channels adjacent to active ones may be misinterpreted as occupied due to interference.
Such sensing errors can lead to incorrect resource block (RB) selection in CR network{s}, resulting in communication failure.
As a result, efficient resource allocation is a complex task in underwater environments.

Additionally, ICI causes differences in the measured interference power at the transmitter and receiver, which can lead to rate or modulation mismatches.
Such differences in the channel introduce inaccuracies in the measured signal-to-noise ratio (SNR).
When using SNR as the channel quality indicator (CQI), this leads to overestimated CQI, making it unreliable when determining the optimal data rate.
To compensate for the defect, channel usage information obtained from spectrum sensing can help estimate the achievable rate of an UWA channel, as it can be used to estimate the interference level.
{Here, SNR-based CQI can be used as a baseline, providing information about the data rate in an ICI-free environment, hence helping to estimate the rate in an asynchronous environment.}
Although various rate control methods have been proposed to address the problem \cite{9736963}, {these} approaches fail to jointly consider the interdependencies between ICI and CQI, which could {enhance performance}.

In summary, CR-based {orthogonal frequency division multiple access (OFDMA)} UWA system suffers from the following challenges:
1) Inaccurate spectrum sensing caused by excess out-of-band emission, and
2) CQI distortion from adjacent channels' ICI, which could lead to rate mismatches.
Therefore, to achieve reliable and fast communications in asynchronous underwater environment, accurate spectrum sensing, precise transmission, and efficient RB allocation are required.

Previous studies on UWA communication have typically focused on either spectrum sensing \cite{10266670, 10168936} or transmission optimization \cite{huang2020adaptive, 9770090}.
In spectrum sensing, energy detection (ED)-based schemes \cite{5703204} offer low sensing time but do not provide {a} resource optimization solution.
Also, {the} implementation of non-coherent OFDMA {in} UWA systems have been studied \cite{9761198, 9609854}, demonstrating resilience against time-varying environments.
However, none of the {existing} works have tackled the challenges of asynchrony by jointly considering spectrum sensing and resource allocation, leaving the topic as an open research gap.
This paper aims to fill the gap by jointly utilizing and optimizing key factors such as UWA spectrum data and transmission rate, which is critical for {an} effective deployment of UWA communication techniques.

\subsection{Proposed Method}  \label{subsec:proposed_method}
In this paper, we propose a deep reinforcement learning (DRL)-based approach for spectrum sensing and resource optimization in asynchronous OFDMA-based UWA-CR networks.
We propose a scheme that enhances spectrum sensing accuracy, which can mitigate competition and collision in multi-user CR networks.
The proposed scheme employs a learning-based framework to reduce errors caused by ICI in spectrum sensing, selecting the available RB while adjusting transmission rates using CQI and the estimated RB usage information.
This approach enables the selection of the most stable and efficient resources for optimal transmission.
Simulation results show that the proposed scheme outperforms baseline methods by {achieving} 42.9 \% better {spectral efficiency (SE)} and a 4.4 \% higher communication success rate.

\section{System Model}  \label{sec:system_model}
\subsection{UWA Environment}
\ifonecol
\begin{figure}
    \centering
    \includegraphics[width=0.7\linewidth]{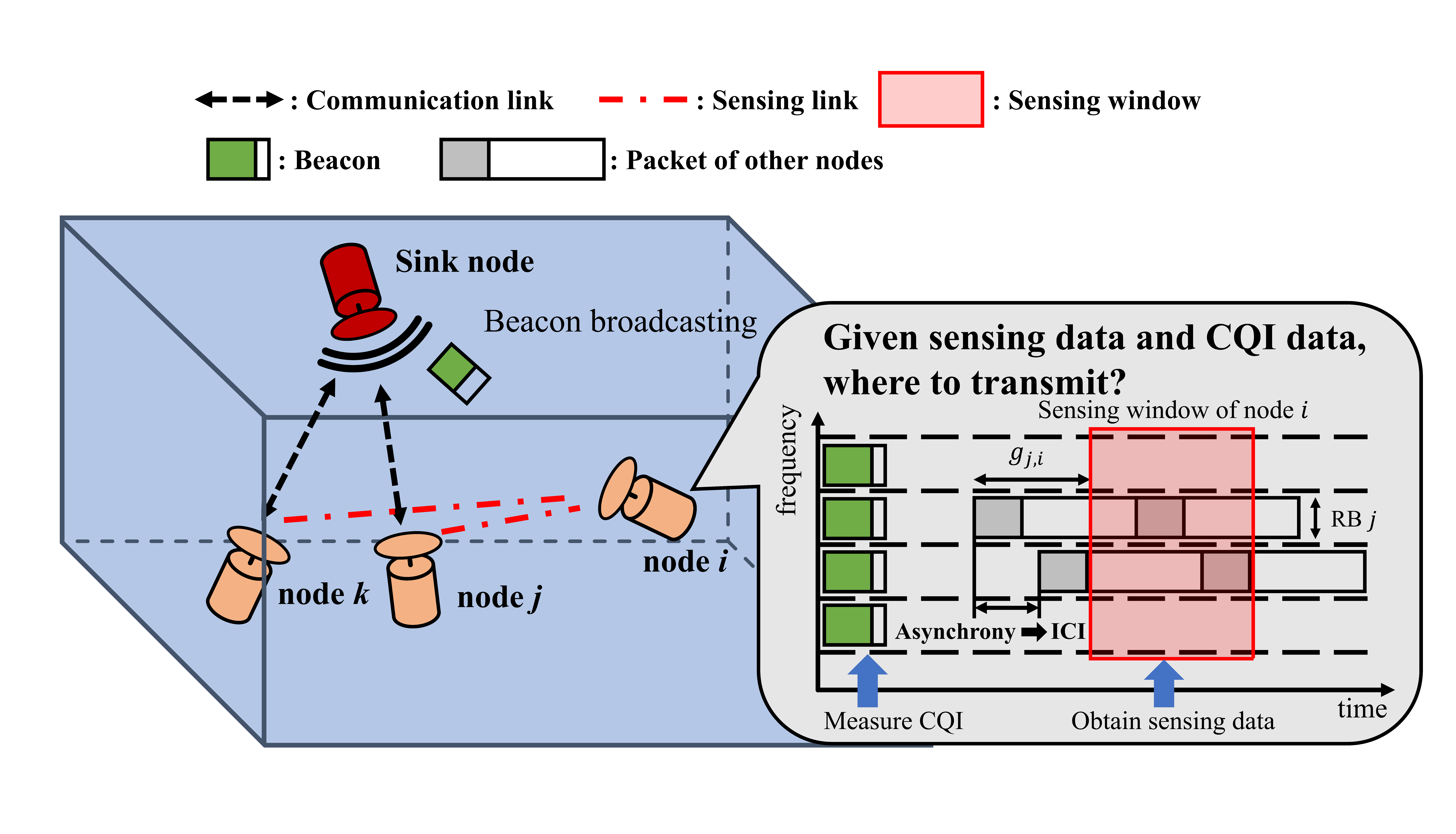}
    \caption{System model of an OFDMA-based UWA-CR network environment.}
    \label{fig:system_model}
\end{figure}
\fi

Figure \ref{fig:system_model} depicts a UWA communication network with OFDMA-based CR protocol, comprising a single sink node and multiple sensor nodes denoted as $\mathcal{S}=\{1, 2, \cdots, S\}$, with each sensor node communicating directly with the sink node.
We represent the number of frequency resources as $N_{\text{RB}}$.
{In accordance with} CR protocols, each sensor node performs spectrum sensing to identify available communication resources.
Spectrum sensing enables a sensor node to detect signals {of} other active sensor nodes; the link between a sensor node and the active node is denoted as the sensing link.

The sink node periodically transmits a beacon signal to all channels, which contains a known sequence.
Sensor node uses the obtained information to determine the downlink CQI, which is then used to infer the uplink CQI using channel reciprocity.
As the beacon transmit period is known to all receivers, we can assume that the transmission collision between the sensor nodes and the sink node does not occur.

\subsection{Data Generation}  \label{subsec:data_generation}

To accurately {model} real underwater channel characteristics, we implemented a UWA channel simulator using {the }BELLHOP Acoustic Toolbox \cite{bellhop_user_guide}, a {widely used} tool for UWA beam tracing and channel modeling.
The tool allows for a more accurate simulation of UWA communications environment, surpassing the limitations of basic transmission loss models.
We generated the {channel impulse responses (CIRs)} of UWA channels and applied the {CIRs} asynchronously to OFDM symbols to obtain the received signal.

To implement a rate mismatch scenario and calculate the correct achievable rates, we derive the signal-to-interference-plus-noise ratio (SINR) of asynchronous OFDMA signals.
We first formulate the signal of node $s$ received by sink node as
\begin{equation}
    y = h_{s}(t) * x_{s}(t) + \sum_{s' \in \mathcal{S} \setminus \{s\}} h_{s'}(t) * x_{s'}(t + g_{s',s}) + z(t),
\end{equation}
where $g_{s',s}$ is the operating time gap between symbols of $s$ and $s'$ as depicted in Fig. \ref{fig:system_model}.
{This random time gap $g_{s',s}$ also helps us implement dynamic UWA environment, with distances between the nodes time-varying.}
Also, $z(t)$ denotes the additive white Gaussian noise (AWGN) with variance $\sigma^2_n$, and $h_s(t)$ denotes the {CIR} of node $s$, obtained using BELLHOP.
Sampling the signal with sampling period $T_{\text{samp}}$, we obtain the discrete time samples of $y(t)$ as
\begin{gather}
    y[n] = h_{s} [n] \circledast x_{s} [n] + J[n] + z[n],
\end{gather}
where $J[n] = \sum_{s' \in \mathcal{S} \setminus \{s\}} h_{s'}(n T_{\text{samp}}) * x_{s'}(n T_{\text{samp}} + g_{s',s})$.

{Following the notations of \cite{9736963},} assuming that the FFT size $N_{\text{FFT}}$ is equal to the number of samples, the frequency-domain received signal's $w$-th OFDM symbol is expressed as
\begin{align}
    \mathbf{y}^{(w)} &= \mathbf{H}_{s}\mathbf{x}_{s}^{(w)} + \mathbf{J}^{(w)} + \mathbf{z} \\
    &= \underbrace{\mathbf{H}_{s} \mathbf{F}^\text{H} \mathbf{d}_{s}^{(w)}}_{\text{{(a) Desired signal}}} + \underbrace{\sum_{s'\in \mathcal{S}_{s}^{(w)}} \mathbf{H}_{s'} \mathbf{x}_{s'}^{(w)}}_{\text{{(b) ICI}}} + \mathbf{z},  \label{eq:matrix_form_signal}
\end{align}
where $\mathbf{H}_{s}$ is a circulant matrix of size $N_{\text{FFT}}\times N_{\text{FFT}}$ constructed from $h_s$, $\mathbf{F}$ is the fast Fourier transform matrix, $\mathbf{d}_s^{(w)}$ is the data vector of node $s$'s $w$-th OFDM symbol, containing data values on the indices corresponding to the allocated subcarriers, and $\mathcal{S}_{s}^{(w)}$ is the set of nodes that can interfere with the $w$-th symbol of node $s$.
Finally, $\mathbf{J}^{(w)} = \sum_{s'\in\mathcal{S}^{(w)}_s} \mathbf{H}_{s'} \mathbf{x}_{s'}^{(w)}$ of equation (4b) is the ICI stemming from asynchrony, where $\mathbf{H}_{s'}$ is a toeplitz matrix constructed from $h_{s'}$, and $x_{s'}^{(w)}$ is the asynchronous OFDM data symbols of node $s'$, windowed to match the OFDM symbol of node $s$, as illustrated in Fig. \ref{fig:system_model}.
{As SNR-based CQI does not consider interference, the CQI {becomes} distorted {based on} the interference power.}

After channel equalization, we obtain
\begin{equation}
    \mathbf{D}^{-1} \mathbf{F} \mathbf{y}^{(w)} = \mathbf{d}_{s}^{(w)} + \mathbf{F} \mathbf{H}^{-1} \mathbf{J}^{(w)} + \mathbf{D}^{-1} \mathbf{F} \mathbf{z},
\end{equation}
where $\mathbf{D}^{-1} = (\mathbf{F} \mathbf{H}_{s} \mathbf{F}^\text{H})^{-1}$.
We formulate the SINR of a subcarrier allocated at $k$-th element of node $s$'s data $\mathbf{d}_{s}^{(w)}$ as
\begin{equation}
    {\fontfamily{cmss} \selectfont \text{SINR}}_{k}^{(w)}
    = \frac{\mathbb{E}[|\mathbf{d}_{s}^{(w)}[k]|^2]}{\mathbb{E}[|(\mathbf{F} \mathbf{H}^{-1} \mathbf{J})[k]|^2] + \sigma^2_n|\mathbf{D}^{-1}(\mathbf{D}^{-1}) ^{\text{H}}|_{k,k}}. \label{eq:sinr}
\end{equation}

Finally, denoting the set of subcarriers allocated to node $s$ as $\mathcal{K}_s$ and the number of OFDM symbols in a packet as $W$, we use the calculated SINR to obtain the actual achievable rate of a packet of node $s$ as ${\fontfamily{cmss} \selectfont \text{rate}}_s=\frac{1}{W|\mathcal{K}_s|} \sum_{w\in W} \sum_{k\in\mathcal{K}_s} \log_2(1 + {\fontfamily{cmss} \selectfont \text{SINR}_k}^{(w)})$.
Similarly, CQI can be calculated as ${\fontfamily{cmss} \selectfont \text{CQI}}
= \mathbb{E}_{\{\mathcal{K}_b\}}[\frac{\mathbb{E}[|\mathbf{d}[k]|^2]}{\sigma^2_n|\mathbf{D}^{-1}(\mathbf{D}^{-1}) ^{\text{H}}|_{k,k}}]$, where $\mathcal{K}_b$ is the set of subcarriers used by the beacon.

\section{Spectrum Allocation based on Deep Reinforcement Learning}  \label{sec:drl}
The complex relationships between ICI, multipath effects, and {a} dynamic environment make heuristic optimization methods ineffective.
Additionally, the multi-faceted objectives of the target task make it challenging to optimize using heuristic methods.
DRL, however, can adapt to non-linear dependencies, enabling faster and efficient resource allocation.

\ifonecol
\begin{figure}
    \centering
    \includegraphics[width=0.8\linewidth]{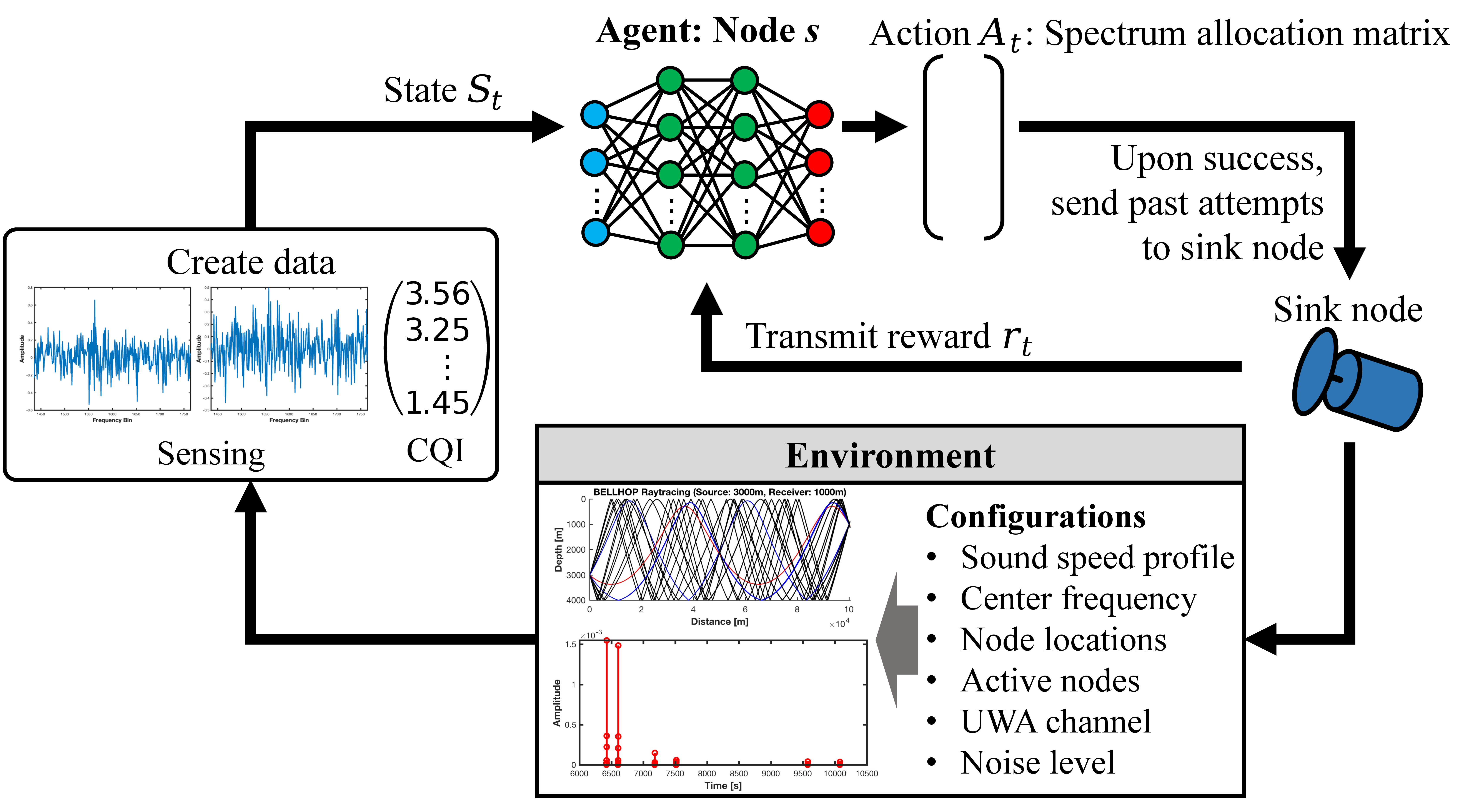}
    \caption{Overview of the interactions between the RL components.}
    \label{fig:rl_system_model}
\end{figure}
\fi

\subsection{Deep Reinforcement Learning Components}  \label{subsec:drl_components}

\subsubsection{State}  \label{subsubsec:state}
{The underwater nodes are the DRL agents of the environment.}
The DRL agent observes the UWA environment and obtains the CQI and UWA spectrum sensing data, denoted as $S_t$ at time step $t$.
SNR is calculated using the beacon from the sink node, and is used as the CQI of the environment.
The CQI data is a vector of length $N_{\text{RB}}$, and is fed directly to the agent.
{UWA} spectrum sensing data is truncated around the signal center frequency $f_c$ and is divided into the real and imaginary parts, creating a matrix of shape $N_{\text{sens}} \times 2$, where $N_{\text{sens}}$ is the number of windowed samples.

\subsubsection{Agent}  \label{subsubsec:agent}
We {use} the Advantage Actor-Critic (A2C) algorithm to {improve} training efficiency.
In A2C, the actor network learns the policy $\pi_{\theta}$ that maximizes the reward, with the help of an advantage function.
The critic network $V_{\mathbf{w}}$ learns to approximate the value of each state represented using a state-value function.
The main network, consisting of the actor and critic networks, is modularized into two parts.
First, the sensing module takes the sensing data as the input and extracts features from the data.
Next, both the CQI data and the filtered sensing data pass through the main module, where the features are combined to process interdependent UWA factors.

\subsubsection{Action}  \label{subsubsec:action}
The action of the agent at time step $t$ is denoted as $A_t$ which is a matrix of dimensions $N_{\text{RB}} \times 2$; the first column vector $\hat{\bm v}_{\text{RB}} \in \mathbb{R}^{N_{\text{RB}}\times 1}$ {represents} the model's RB selection, while the second column vector $\hat{\bm v}_{\text{rate}}\in \mathbb{R}^{N_{\text{RB}}\times 1}$ specifies the {estimated transmission rates}.
Here, $\hat{\bm v}_{\text{RB}}$ represents a probability distribution where the agent estimates the RB with the highest likelihood of availability.
We calculate the element-wise product of two vectors to obtain the chosen RB index as $\argmax (\hat{\bm v}_{\text{rate}} \cdot \hat{\bm v}_{\text{RB}})$ with highest expected throughput, and the chosen rate is $\hat{{\bm v}}_{\text{rate}}[\argmax (\hat{\bm v}_{\text{rate}} \cdot \hat{\bm v}_{\text{RB}})]$.

\subsubsection{Reward}  \label{subsubsec:reward}
{Using throughput as the reward,} the agent may achieve a high throughput by selecting an extremely high transmission rate in the expense of lower success rate.
Such behavior is suboptimal, as frequent communication attempts {increase} communication overhead.
To avoid such behavior, we introduce two {auxiliary terms: $r_{\text{RB}}$ and $r_{\text{rate}}$.}

We define the RB selection reward $r_{\text{RB}}$ as a negative weighted sum of the cross-entropy (CE) between the RB estimate $\hat{\bm v}_{\text{RB}}$ and the ideal RB index $\argmax ({\bm v}_{\text{rate}}) \in \{1, \cdots, N_{\text{RB}}\}$, and the mean squared error (MSE) between $\hat{\bm v}_{\text{RB}}$ and available RBs ${\bm v}_{\text{RB}} \in \mathbb{R}^{N_{\text{RB}}\times1}$.
We construct a vector with actual achievable rates ${\bm v}_{\text{rate}}$ constructed from ${\fontfamily{cmss} \selectfont \text{rate}}_s$, with rate of occupied RB denoted as 0.
Also, ${\bm v}_{\text{RB}}$ is a binary vector where available RB is denoted as 1.
The formulated reward is
\begin{align}
    r_{\text{RB}} = & - w_1 \cdot CE(\hat{\bm v}_{\text{RB}}, \argmax ({\bm v}_{\text{rate}})) \nonumber \\
    & - w_2 \cdot MSE(\hat{\bm v}_{\text{RB}}, {\bm v}_{\text{RB}}),  \label{eq:rb_reward}
\end{align}
where $CE(x, y)$ is the CE loss between $x$ and $y$ and $MSE(x, y)$ is the MSE loss between $x$ and $y$.

For rate selection reward $r_{\text{rate}}$, we calculate the MSE between estimated rates $\hat{\bm v}_{\text{rate}}$ and actual achievable rates ${\bm v}_{\text{rate}}$ across all RBs, setting any values of $\hat{\bm v}_{\text{rate}}$ exceeding ${\bm v}_{\text{rate}}$ to zero to penalize rate mismatch.
Also, the MSE loss between $\hat{\bm v}_{\text{rate}}[\argmax ({\bm v}_{\text{rate}})]$ and $\max({\bm v}_{\text{rate}})$ is calculated.
Calculating the negative weighted sum of the terms, $r_{\text{rate}}$ is obtained as
\begin{align}
    r_{\text{rate}} = & - w_3 \cdot MSE(\hat{\bm v}_{\text{rate}}, {\bm v}_{\text{rate}}) \nonumber \\
    & - w_4  \cdot MSE(\hat{\bm v}_{\text{rate}}[\argmax ({\bm v}_{\text{rate}})], \max({\bm v}_{\text{rate}})).  \label{eq:rate_reward}
\end{align}
{Using the extra terms and throughput, the reward $r_t$ {at} time step $t$ is {given as}:}
\begin{equation}
    r_t = w_5 \cdot r_{\text{RB}} + w_6 \cdot r_{\text{rate}} + w_7 \cdot \hat{\bm v}_{\text{rate}}[\argmax (\hat{\bm v}_{\text{rate}} \cdot \hat{\bm v}_{\text{RB}})].  \label{eq:reward}
\end{equation}

When the node connects successfully with the sink node, previous reports on transmission attempts are sent to the sink node as illustrated in Fig. \ref{fig:rl_system_model}.
As the sink node knows all information about the RB usage and the SINR of the nodes, calculation of the transmission rate and the reward is possible.

\subsection{Training Procedure}  \label{subsec:training_procedure}

\begin{algorithm}[t]
\caption{A2C Agent Training Procedure} \label{alg:training_procedure}
\begin{algorithmic}[1]
\State \textbf{Initialize:} Actor network $\pi_\theta$, critic network $V_{\mathbf{w}}$ and set {episode} counter $T=0$
\Repeat  \Comment{\textit{Generate episodes}}
    \State Randomly initialize node locations
    \State Randomly select active nodes
    \State Set time step counter $t \gets 0$
    \Repeat
        \State Generate channels using BELLHOP
        \State {Agent reads the spectrum and calculates the SNR}
        \State $A_t \sim \pi_{\theta}$  \Comment{\textit{Actor chooses action}}
        \State Update state $S_t$, get reward $r_t$ from $V_\mathbf{w}$
        \State Store $(S_t, A_t, r_t)$
        \State $t \gets t + 1$
    \Until{Episode termination}
    \State Calculate GAE $a_t^{\text{GAE}(\gamma, \lambda)}$
    \State Update actor network $\pi_{\theta}$ and critic network $V_{\mathbf{w}}$
    \State $T \gets T + 1$
\Until{$T\geq T_{\text{max}}$ or $\theta$ is good}  \Comment{\textit{Monitor reward}}
\end{algorithmic}
\end{algorithm}

The DRL agent training process is depicted in Algorithm \ref{alg:training_procedure}.
At the start of each episode, the network and the environment are initialized.
{Beacon is used to calculate the SNR, and CIR is used to calculate the UWA spectrum data.}
The state is then fed to the agent, which generates the corresponding action.
Finally, the reward is calculated and is returned to the agent.
At the end of each episode, the accumulated rewards are used to update the agent.
In actual deployment, online learning is executed with the reward received from the sink node, which allows the agent to adjust to the environment.

More specifically, as simulations do not fully mimic the actual UWA channel characteristics, the agent needs to learn new features unique to the real-world.
As training volume is crucial in deep learning-based approaches, we use generalized advantage estimation (GAE) \cite{DBLP:journals/corr/SchulmanMLJA15} to boost stability in dynamic UWA environments, improve sample efficiency, and accelerate convergence.
Such properties of GAE also allow effective training in simulation, as training a DRL agent in time-varying UWA environment requires substantial training samples.
Given reward $r_t$ and state $S_t$, the $n$-th order advantage function that helps evaluate the given action is described as
\begin{equation}
    a^{(n)}_t=\sum_{i=1}^{n} \gamma^{i-1} r_{t+i} + \gamma^n V(S_{t+n})-V(S_t),  \label{eq:advantage}
\end{equation}
where $V(S_t)$ is the estimated value of $S_t$, and $\gamma$ is the discount factor.
Generalizing the above result, we obtain GAE as 
\begin{equation}
    a_t^{\text{GAE}(\gamma, \lambda)}=(1-\lambda)\sum_{n}\lambda^{n-1} a_t^{(n)},  \label{eq:gae}
\end{equation}
where $\lambda$ is the bias-variance tradeoff parameter.

{
\textbf{Convergence Analysis.}
As the proposed algorithm has bounded parameters, has Lipschitz continuous GAE, uses asynchronous time scales to update the actor and critic, and the feature space of critic spans the feature space of actor, the algorithm converges given judicious learning rates \cite{NIPS1999_6449f44a}.
}

\section{Simulation Results}  \label{sec:simulation_results}
\subsection{Simulation Settings}  \label{subsec:simulation_settings}

\ifonecol
    \begin{table}[]
    \centering
    {\tiny
    \centering
    \caption{Environment Parameters and DL Model Configurations}
    \adjustbox{width=0.5\linewidth}{
    \begin{tabular}{@{}cc@{}}
    \toprule
    \textbf{Parameters} & \textbf{Values} \\
    \midrule
    Underwater space & 1000 $\times$ 1000 $\times$ 200 (m) \\
    OFDM symbol duration & 128 (ms) \\
    CP duration & 30 (ms) \\
    Center frequency ($f_c$) & 1.2 (kHz) \\
    $T_{\text{samp}}$ & 45.455 ($\mu$s) \\
    $N_{\text{FFT}}$ & 256 \\
    Number of RBs & 5 \\
    Number of subcarriers per RB & 10 \\
    \bottomrule
    \end{tabular}
    }
    \label{tab:simulation_settings}
    }
    \end{table}
\fi

The OFDM symbol and waveform parameters are listed in Table \ref{tab:simulation_settings}.
{Delay spread measurements of \cite{5290161, 4686807} suggest that the appropriate cyclic prefix (CP) length should be larger than 30 ms and the OFDM symbol length should exceed 100 ms.
We select 128 ms as the OFDM symbol duration \cite{8924527}.}
We use 1.2 kHz as the $f_c$ although a higher frequency is more commonly used.
This is because the low $f_c$ allows for longer communication range \cite{DOMINGO2008163, 4752682}, where the problem of interest is more predominant.
The number of RBs is set to $N_{\text{RB}}=5$.

We implement five baseline methods.
To compare the DRL agent with other deep learning (DL) model architectures, we include a convolutional neural network (CNN)-based model, since it has {been proven} to perform well in asynchronous spectrum sensing tasks \cite{10168936, app12094534}.
The CNN model size is matched to that of the DRL agent, as CNN models are known for {their} good performance given similar memory size.
More specifically, the DRL agent size is approximately 84.3 MB, which is slightly less than the CNN model size of 88.3 MB.
A fully connected network (FCN) architecture is also used.

Additionally, we employ three heuristic methods.
In CQI-$\epsilon$ methods, RB with $k$-th best CQI is chosen, and the transmission rate is determined according to the Shannon capacity.
To compensate for the rate distortion stemming from interference, we use $\epsilon$-rate control method where the transmission rate is lowered by a factor of $\epsilon$.
Therefore, the actual transmission rate is calculated as ${\fontfamily{cmss} \selectfont \text{rate}}_{\text{transmit}} = (1-\epsilon) {\fontfamily{cmss} \selectfont \text{rate}}_{\text{CQI}}$, where ${\fontfamily{cmss} \selectfont \text{rate}}_{\text{CQI}}$ is the rate calculated based on the CQI.
We choose $\epsilon=0.37$, as prior experiments showed that the value guaranteed a 90 \% rate decision success rate.
For CQI-$\epsilon$ methods, we use $k=1$ and random $k$ value, where RB is chosen in random.
In ED-$\epsilon$ method, RB is chosen according to the RB energy, which is measured using ED method.
Next, transmission rate is chosen according to $\epsilon$-rate control method.
We use $\epsilon=0.39$, which showed the best sum {SE}.

\subsection{Performance Evaluation}  \label{subsec:performance_evaluation}

\ifonecol
\begin{figure}
    \centering
    \includegraphics[width=0.6\linewidth]{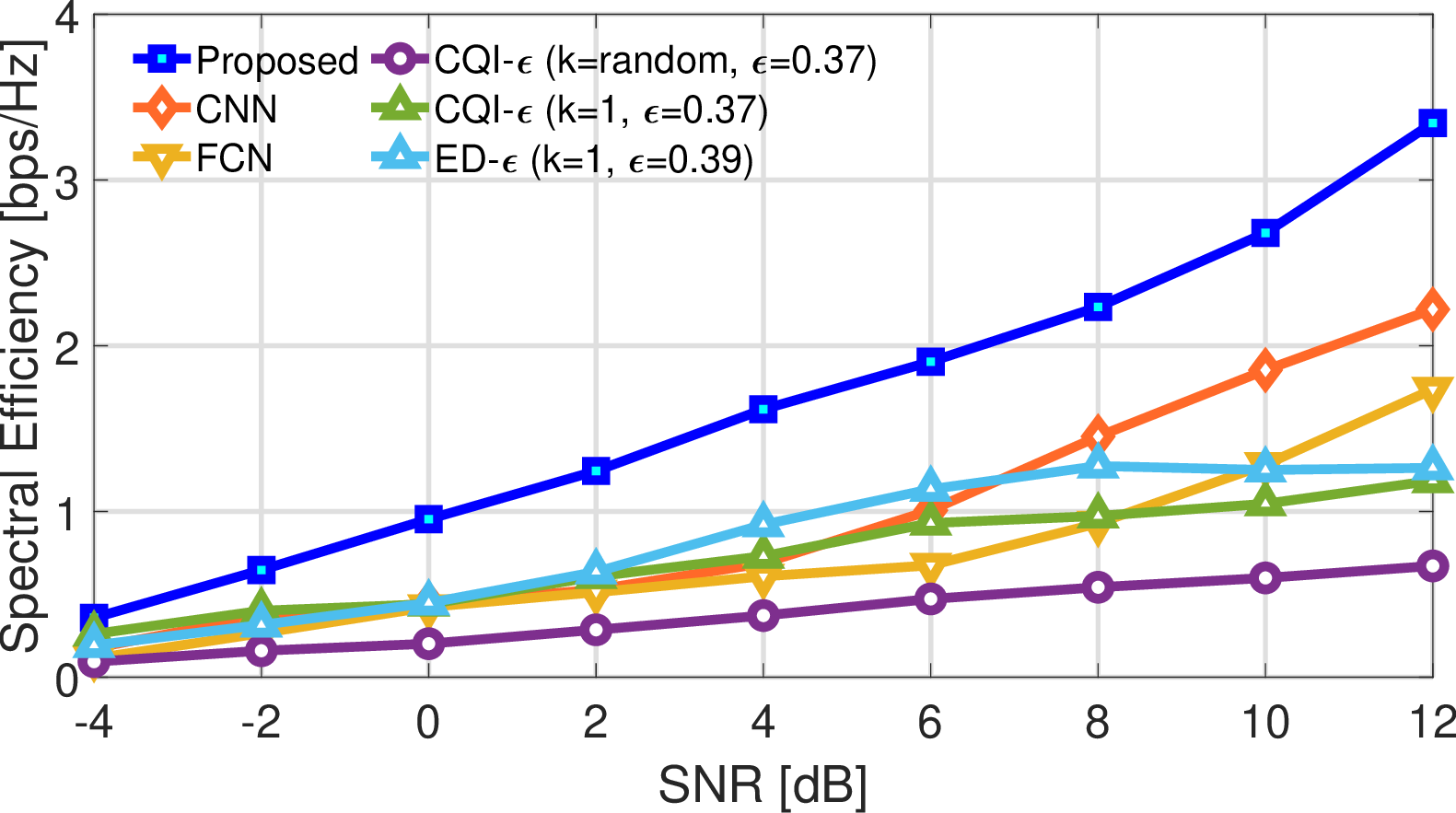}
    \caption{Average network SE of different schemes for different SNR values.}
    \label{fig:throughput}
\end{figure}
\fi

\ifonecol
    \begin{table}[]
    \centering
    \caption{Transmission Rate and Success Rate of the Schemes in an Environment with an SNR of 6 \textnormal{dB}}
    \adjustbox{width=0.7\linewidth}{
    \begin{tabular}{@{}ccccccc@{}}
    \toprule
    \textbf{Schemes} & \textbf{Proposed} & \textbf{CNN} & \textbf{FCN} & \multicolumn{1}{c}{\begin{tabular}[c]{@{}c@{}}\textbf{CQI-$\epsilon$}\\ \textbf{($k$=random)}\end{tabular}} & \multicolumn{1}{c}{\begin{tabular}[c]{@{}c@{}}\textbf{CQI-$\epsilon$}\\ \textbf{($k$=1)}\end{tabular}} & \multicolumn{1}{c}{\begin{tabular}[c]{@{}c@{}}\textbf{ED-$\epsilon$}\\ \textbf{($k$=1)}\end{tabular}} \\ \midrule
    SE {[}bps/Hz{]} & 1.90 & 1.01 & 0.67 & 0.47 & 0.93 & {1.33} \\
    Success rate {[}\%{]} & 65.8 & 42.5 & 59.5 & 38.9 & 48.4 & {63.0} \\ \bottomrule
    \end{tabular}
    }
    \label{tab:scheme_stats}
    \end{table}
\fi

{Each method is evaluated in a UWA environment with an SNR range of -4 to 12 dB, a range commonly used in other studies \cite{10168936, huang2020adaptive, 9770090}.}
SE is calculated as the average transmission rate per Hz, assigning 0 bps/Hz for failed transmissions.
Communication is regarded as a success if the selected transmission rate does not exceed ${\fontfamily{cmss} \selectfont \text{rate}}_s$, and the chosen RB is unoccupied.

From Fig. \ref{fig:throughput}, the proposed A2C agent surpasses all baseline methods in SE.
More specifically, at an SNR of 10 dB, the proposed scheme reaches 2.68 bps/Hz, outperforming the best baseline method with 1.85 bps/Hz by 49.9 \%.

Table \ref{tab:scheme_stats} provides a detailed comparison of each scheme's SE and success rate.
The proposed method demonstrates superior SE performance, achieving a SE of 1.90 bps/Hz --- approximately 42.9 \% higher than the best baseline method.
Furthermore, the proposed method yields a success rate of 65.8 \%, about 4.4 \% higher than the best baseline model.

The reward curve of Fig. \ref{fig:reward_curve} demonstrates a steady increase as training progresses, while average throughput reaches a peak early in the process.
This behavior is guided by the reward structure in equation \eqref{eq:reward}.
Initially, exploration guides the agent toward a suboptimal solution {with} high transmission rate and low success rate, {leading to a} surge of the throughput and the drop in success rate.
In the next phase, the RB selection reward guides the agent towards better success rate and lower transmission rate, resulting in a higher reward.
This is illustrated from the decline in the throughput graph, alongside the upwards trend of the success rate.
{Next, auxiliary reward term of equation \eqref{eq:rate_reward} further tunes the agent.}
Consequently, the model converges toward an optimal policy that satisfies all constraints of the reward.

\ifonecol
\begin{figure}
    \centering
    \includegraphics[width=0.7\linewidth]{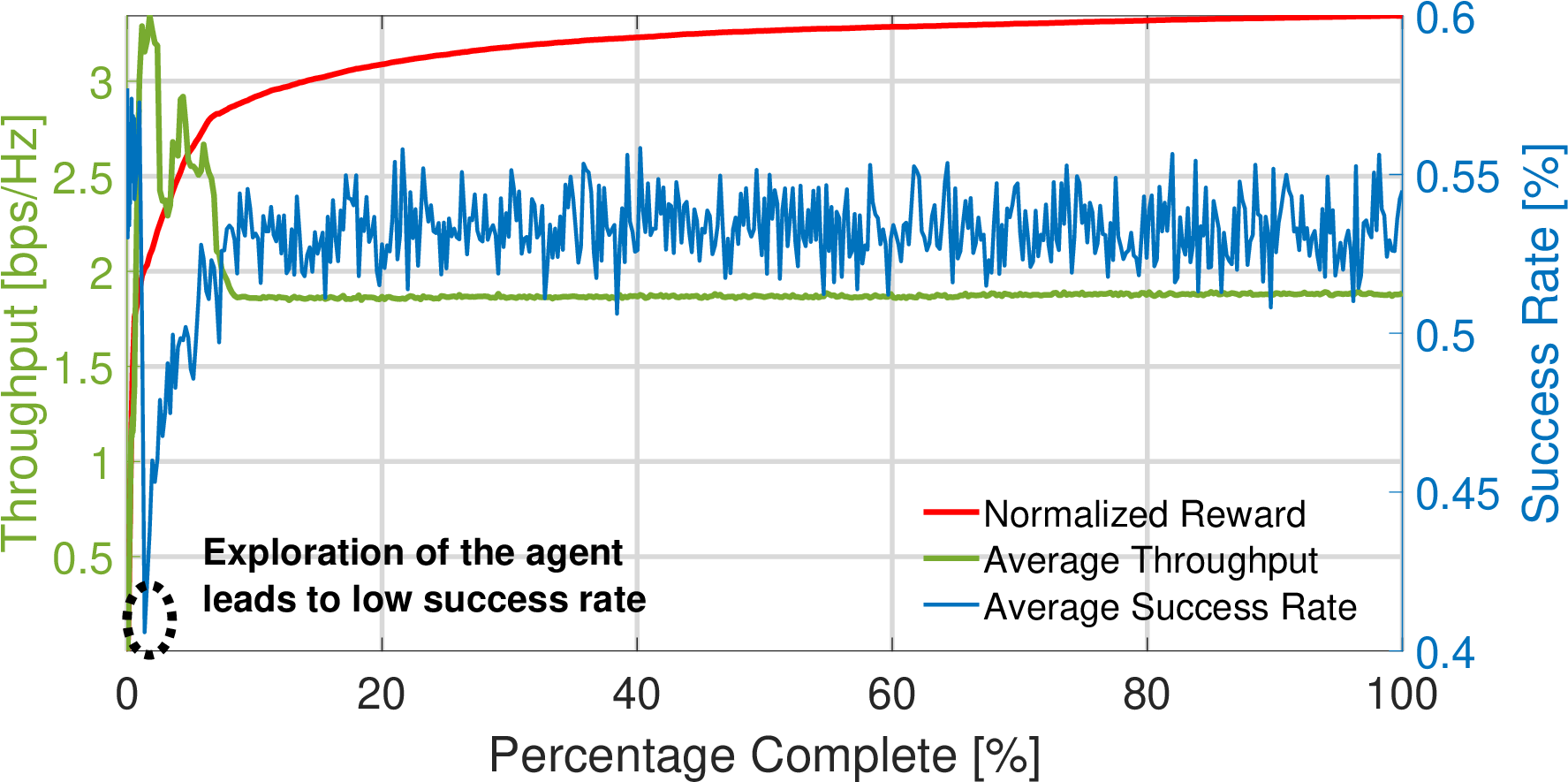}
    \caption{Reward, throughput, and success rate of the agent during training.}
    \label{fig:reward_curve}
\end{figure}
\fi

\section{Conclusion}  \label{sec:conclusion}
This letter proposes an A2C-based spectrum sensing and resource allocation protocol in {a} UWA communication environment.
The proposed protocol addresses asynchrony, {which is} a critical yet overlooked factor in underwater communications research.
The proposed method resolves the complexity arising from interdependent underwater environment factors and successfully demonstrates {its performance}.

Although we assumed good channel reciprocity properties, the assumption may not {hold for long-range} or highly mobile UWA communications.
Future works could enhance the method by {considering mobility}, using other CQI {measurement} methods, and examining the effects of interference and collision in multi-user environment. 

\bibliographystyle{IEEEtran}
\bibliography{refs}

\end{document}